# Bayesian Networks in Healthcare:
## *Distribution by Medical Condition*

Scott McLachlan[1,2], Kudakwashe Dube[2,3], Graham A Hitman[4], Norman Fenton[1], Evangelia Kyrimi[1]

[1] *Risk and Information Management, Queen Mary University of London, United Kingdom*
[2] *Health informatics and Knowledge Engineering Research (HiKER) Group*
[3] *School of Fundamental Sciences, Massey University, New Zealand*
[4] *Centre for Genomics and Child Health, Blizard Institute, Queen Mary University of London, London, United Kingdom*

Abstract

Bayesian networks (BNs) have received increasing research attention that is not matched by adoption in practice and yet have potential to significantly benefit healthcare. Hitherto, research works have not investigated the types of medical conditions being modelled with BNs, nor whether any differences exist in how and why they are applied to different conditions. This research seeks to identify and quantify the range of medical conditions for which healthcare-related BN models have been proposed, and the differences in approach between the most common medical conditions to which they have been applied. We found that almost two-thirds of all healthcare BNs are focused on four conditions: cardiac, cancer, psychological and lung disorders. We believe that a lack of understanding regarding how BNs work and what they are capable of exists, and that it is only with greater understanding and promotion that we may ever realise the full potential of BNs to effect positive change in daily healthcare practice.

## 1. Introduction

Bayesian Networks (BNs) have received increasing attention during the last two decades [1, 2] for their particular ability to be applied to challenging issues and aid those making decisions to reason about cause and outcome under conditions of uncertainty [3-5]. In 2016, the journal *Machine Learning* ran a special issue on *Machine Learning for Healthcare and Medicine* [6]. In explaining their motivation for that issue, the editors discussed how ever-increasing volumes of health data created potential for developing new knowledge that could improve the practice of patient care. Machine learning (ML) and artificial intelligence (AI) approaches have been proposed for a diverse range of health topics from genomics [7, 8] to treatment selection [9, 10] and outcome, prognosis, prediction [11]. A significant benefit of BNs over methods of pure ML from data is that BNs do not explicitly require massively large datasets. BNs can incorporate the accumulated knowledge of experts in circumstances where data are limited, and still produce meaningful and accurate decision-support systems. This paper is part of a larger effort to address the wide chasm between research enthusiasm for BNs and their lack of adoption in healthcare. The paper seeks to understand the medical conditions BNs are being used to model and make predictions for, and any apparent differences in their application between clinical domains, in order to shed light on how we may harness the potential of BNs in daily healthcare practice.

BNs are based on Bayes' theorem and use a graphical approach for *compact representation of multivariate probability distributions* and *efficient reasoning under uncertainty* [2]. Bayes' theorem is a formula for understanding how belief in the probability of phenomenon under observation evolves as knowledge of related influential phenomena increases [12, 13]. For instance, where multiple diseases present with similar symptoms making differential diagnosis challenging, clinicians must update their degree of belief about which illness is causing the patient's poor health as new information becomes available from diagnostic tests and physical examination of the patient.

There are three approaches for developing a BN; (1) using only data (data-driven BNs) (2) using only knowledge (expert-driven BNs), and (3) using a combination of both data and knowledge (hybrid BNs). Expert-driven and hybrid BNs, which supplement data with knowledge, could potentially be the most capable approach for supporting Learning Health Systems (LHS), precision medicine, and thus enabling personalised clinical decision-making from large collections of aggregated health data. LHS, which include such well-known types as clinical decision support systems (CDSS) [14], have experienced slow adoption and are not routinely observed in clinical practice [15-17]. This is particularly true for those LHS based on BNs [2]. While others have reviewed the scope for AI and ML in healthcare [18-20], in each case BNs have been overlooked.

Authors who have reviewed medical decision support models tend to provide a *brief but broad brushstrokes* view of either the ML or AI domain before focusing the bulk of their work on whichever method was their particular area of research interest. An example is [21] where the authors provide non-exhaustive lists and a brief glossary of domain-wide terms before focusing solely on learning approaches for neural networks, which are only one of a vast number of AI and ML types. In [18] the authors describe a list of ML algorithms but only expound on two from their research: support vector machines and neural networks. Attempts had been made to identify [19, 22] and even quantify [18] the broad scope of medical conditions and clinical questions or activities to which ML and AI respectively are being applied. However, *it was difficult from each author's method and descriptions to identify: (a) the search terms they had used; (b) a list of the medical conditions or clinical questions; or (c) an accurate accounting of the papers identified by and used in each study.*

The aim of this paper is to establish the medical conditions to which BNs are being applied as part of our wider effort [23] to determine the potential benefits and challenges being faced by those seeking to integrate BNs into daily practice. This paper seeks to achieve its aim by identifying and quantifying the range of medical conditions for which the healthcare-related BN models are being considered. This paper contributes to our overall understanding of the purposes for which BNs are being considered as a predictive or decision-making tool in the healthcare domain, and the differences in approach between the most popular medical conditions to which they have been applied. It is only through continued research to understand the challenges and issues that we may in future harness the potential of BNs to affect positive change, and improve patient outcomes.

This paper is organised as follows: Section 3 presents the method used to achieve the objectives of this paper as part of our larger scoping review. Section 4 presents the results, followed by a discussion of the general findings and their significance in Section 5. The paper concludes with a summary in Section 6.

## 2. Method

We used the literature collection from our scoping review on BNs in healthcare [25]. The search term used to derive that collection was:

```
"(((Bayes OR Bayesian) AND network) OR (probabilistic AND graphical AND
                  model)) AND (medical OR clinical)"
```

Terms such as "Bayesian networks" or "graphical probabilistic models" were used here because they are widely used in the targeted literature. Different ways for explaining the medical condition do occur: in some papers the exact condition is mentioned, while in others broader terms such as "medical or clinical application", "medical or clinical condition", or "medical or clinical setting" are used. Our scope review settled on the broader terms "medical" or "clinical" as they were found in a wider collection of papers. Searching for specific medical conditions would have been impractical as there are many thousands of distinct known conditions.

The search initially identified 3810 papers. Through the process displayed in Figure 1 we then exclude papers published before 2013, those not published in English, and any whose primary content was not healthcare-related. In addition, papers that did not provide BNs and instead were focused merely on Bayesian statistics or meta-analyses were excluded. Graphical models, such as *naive BNs*, which are the simplest BNs and structurally assume all variables are independent, and other graphical computational approaches used in ML or AI, including *neural networks*, have also been excluded.

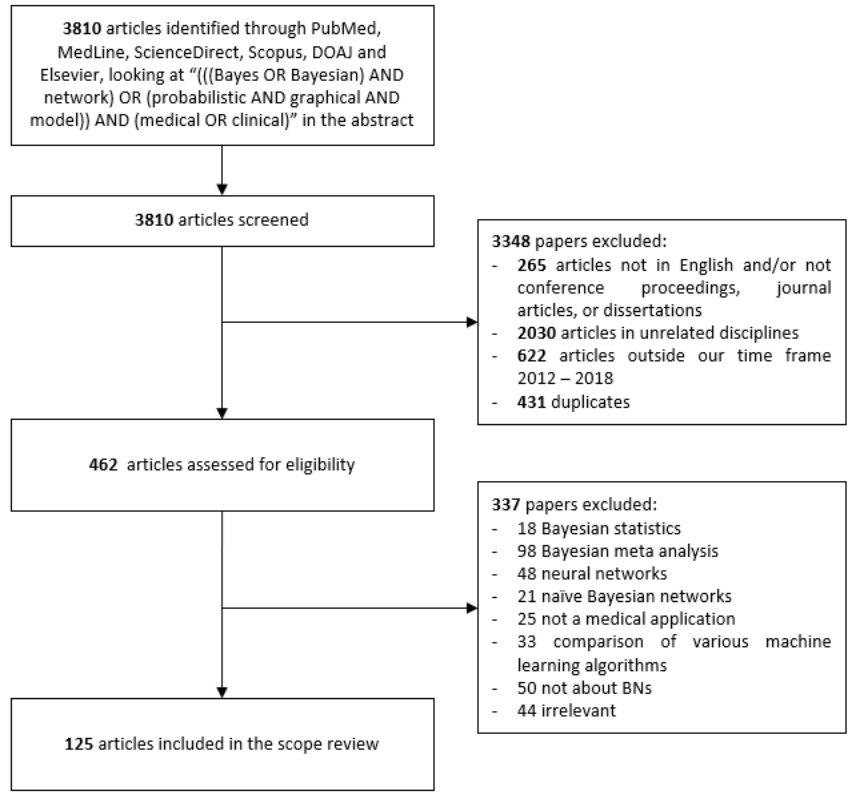

Figure 1: PRISMA diagram (from [23])

    Our primary review plan identified six primary objectives that included identifying the aim of each reported BN, understanding the development processes and inputs used by authors, and how useful the resulting BN may be in clinical practice. The focus of research reported in this paper is Objective 5: to identify distribution and frequency of the medical conditions targeted by the BN models described in the literature collection used in our larger scoping review of BNs in healthcare. Objective 5 is identified in purple in our review concept plan [23] presented here as Figure 2.

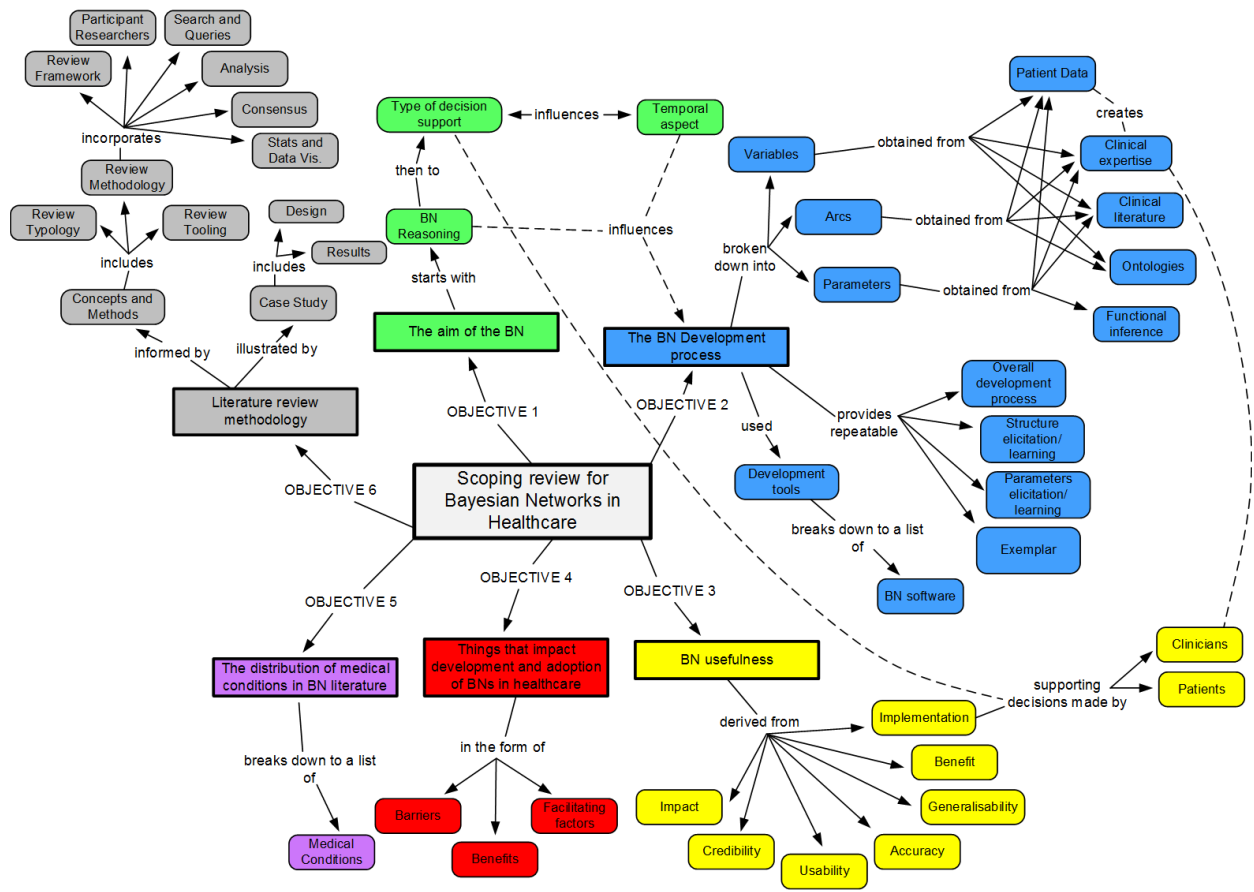

Figure 2: Concept map for the Scoping Review of Bayesian Networks in Healthcare (from [23])

The list of medical conditions in Table 1 was developed with the input of two clinical experts and refined inductively during a small-scale preliminary review described in [2]. Each paper was reviewed by two reviewers who recorded the target medical condition against the list which was presented to reviewers as part of a secure online survey used to manage and conduct the overall review. Where the medical condition was not listed, or the reviewer felt unable to make an explicit classification, it was possible to manually enter details or references to the condition from the literature into a free-text field exposed below the list. Where the two responses for a given paper differed, two authors (SM and EK) reviewed the paper collaboratively to achieve consensus. In rare circumstances where consensus could not be achieved, a clinician was available to assist with classification.

|  | **Medical Condition** | **Exemplar Citations** |
|---|---|---|
| 1 | Blood disorders | [61, 63] |
| 2 | Brain or spine (including CNS injury) | [64] |
| 3 | Chronic Conditions (e.g.: diabetic and arthritic conditions) | [65, 66] |
| 4 | Cancer | [30-32] |
| 5 | Cardiac conditions | [25-27] |
| 6 | Fatigue | [67] |
| 7 | Genetic | [68] |
| 8 | Infectious diseases | [69] |
| 9 | Liver disease | [70, 71] |
| 10 | Lung and breathing disorders | [46-48] |
| 11 | Medication (concentration or reaction to) | [72] |
| 12 | Musculoskeletal Conditions | [73] |
| 13 | Oesophageal disorders (swallowing or speech) | [74] |
| 14 | Organ disorders or failure | [75] |
| 15 | Pregnancy disorders | [76] |
| 16 | Psychological or psychiatric disorders | [41-43] |
| 17 | Skin (burns or disorders) | [77, 78] |
| 18 | Sleep disorders | [79] |
| 19 | Surgery related infection | [80] |
| 20 | Utility of or experience of healthcare | [81] |
| 21 | Unclassified or Other | [52, 82] |

Table 1: list of Medical Conditions

## 3. Results and Discussion

Even though the quantity of BN literature is rapidly increasing, it was possible to identify a distinct focus on a small group of conditions. Specifically, as shown in Figure 3, cardiac conditions, cancer, psychological and psychiatric disorders, and lung and breathing disorders made up 59% of the medical conditions in the literature. This result was not surprising given that other research identified similar disease foci for AI generally [18], and the notoriety of these conditions in the mainstream media as those *most likely to kill you* [24]. The remaining 41% of papers were spread across a diverse collection of seemingly random topics. Figure 3 also illustrates the general distribution of BN literature against the entire list of medical conditions. The complete review dataset is available as a supplemental material spreadsheet on the PamBayesian website at (http://www.pambayesian.org/interim-data/).

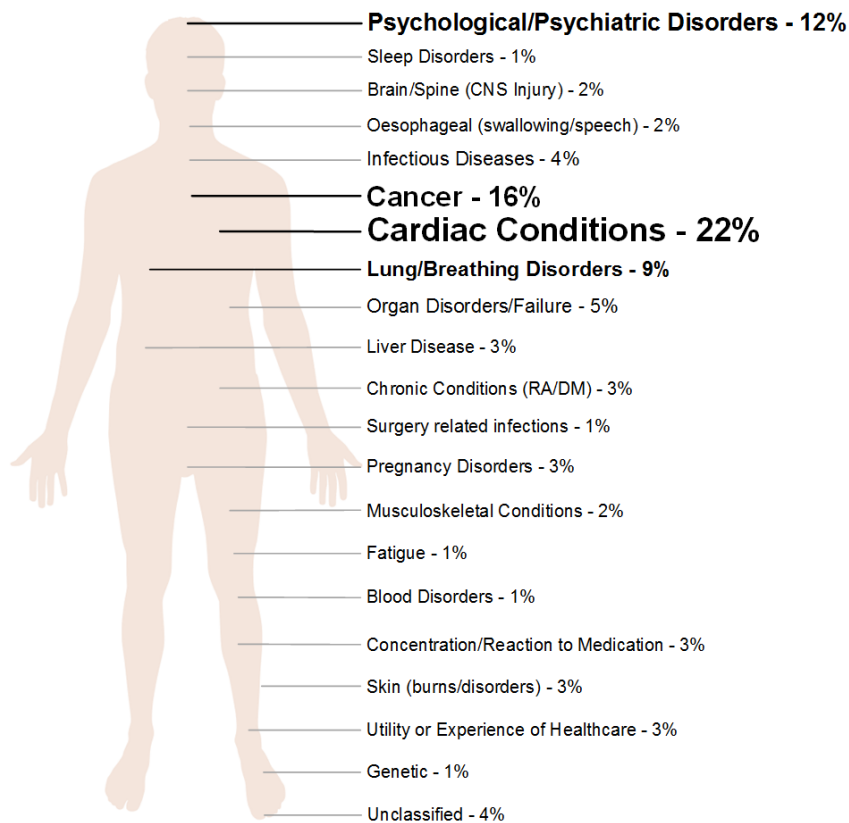

Figure 3: Distribution of medical conditions in BN literature

## 3.1 Most Common Medical Conditions

**Cardiac conditions**

A strong theme for cardiac models was a restriction of focus to either *acute diagnosis* or *prediction of disease progression*. Common to many was the authors' use of electronic patient data as the primary source for elements of BN structure and observations necessary for prediction. This, while simultaneously admonishing the poor quality they observed of most available electronic health record (EHR) data and the limitations that arose from this quality issue [25-27]. Proposed approaches ranged from models for classifying the diagnosis [28] or severity [29] of a patient's condition, to prediction of risk for acute clinical events [25] and the likely progression of the individual's disease [26]. It was not unusual for cardiac BNs to incorporate known comorbidities and validate their interaction with the patient's cardiac condition [26, 27].

**Cancer**

Breast cancer [30-32] and (onco)gene expression [33] separately, and sometimes together [34, 35], were targets of significant interest. BNs have been applied to supporting and enhancing expert knowledge in diagnosing breast cancer from mammography images [36], classifying tumours [30], and as a potential method to improve expert judgement in situations where increasingly complex treatment options have made clinical decisions difficult [37]. Cancer BNs were more likely to seek previously unknown relationships either between symptoms and syndromes [38], or expressed genes and target metastatic cancers [34].

**Psychological and psychiatric disorders**
The most prevalent conditions modelled in the literature were depression [39-41] and the age-related cognitive degeneration diagnoses: dementia and Alzheimer's Disease (AD) [41-43]. Incorporating an organisational structure from ontologies and merging this with BNs was a popular approach in these medical models [39, 44], along with a stronger focus on identifying or relying on a hierarchy of symptoms [39, 44]. While medical models were generally seen to be more likely to draw on experts' knowledge than data alone, the nature and mystery of psychological and psychiatric disorders may be the reason they more frequently drew on clinical expertise during BN development [39-41, 43, 45].

**Lung and breathing disorders**
Half of all models for lung and breathing disorders sought to predict the risk of exacerbation of an already diagnosed chronic clinical condition, including: chronic obstructive pulmonary disease (COPD) [46, 47] and asthma [48]. Others sought to use clinical signs and symptoms to either assess the probability of a particular acute pulmonary diagnosis [49], predict future severity [50], or classify the subtype of the diagnosed disease [51]. One approach proposed a mobile smartphone app that used questionnaires and a BN model to ascertain the patient's current condition and provide contextually relevant advice, while delivering data on the patient's health status to clinicians via an internet connection and a central server [46]. This is similar to the approach proposed by the authors of this paper for their PamBayesian project (www.pambayesian.org).

## 3.2 Researcher and Content Classification

Overall, papers could generally be classified as one of three types. *First,* there were papers written entirely by computer scientists and mathematicians that whilst being dense on technical detail and description for the math of BNs, were sparse on clinical detail for the conditions being modelled - these we describe as *method-driven* [28, 39, 52, 53]. *Second,* there were those that were written entirely by clinicians who included comprehensive discussion and contextualisation of the medical condition and its symptomatology, but significantly less technical discussion of the theory and development of their BN - these we felt were *problem-driven* [54, 55]. *Finally*, there were those presented by a mix of both clinicians and computing or decision scientists - these we describe as *hybrid-driven*.

In *method-driven* works, the focus is mainly on the BN methodology. A medical application is used simply as a case study to evaluate the proposed methodology.

For *problem-driven* works, it was rare to find a paper that whilst appearing to be written entirely by clinicians, was methodological in both its regard for the BN's development process and comprehensive in its technical description of the BN generally. One such unicorn paper drew the attention and comments of both reviewers [56]. The reviewers also generally felt that radiologists came across better than most other clinical disciplines at describing the mathematical and technical aspects of BN development, possibly resulting from their medical physics training [57, 58].

In the *hybrid-driven* works, reviewers tended to introduce both the medical condition (the problem aspects) and BN theory (the methodology aspects) early [48, 59-62]. Having a multidisciplinary team offers the advantage that both aspects receive equal consideration. From the medical perspective, the research problem and benefit of the BN to clinical practice are better explained. From the computing point of view, the overall BN development process is described in greater detail. In papers where both clinicians and computer scientists are involved the BN development and validation process were found to possess greater accuracy, as clinical input was considered.

### 3.3 Strengths and Limitations

A strength of this review is that this is a component of the largest scoping review of BNs in healthcare [23]. Another strength is that tools, including a structured list of medical conditions and an online standardised survey form were developed to ensure a consistent review process.

A limitation of this review is that, even if this is a representative sample of papers published in both medical and AI journals and conference proceedings, it may not reflect the entire range of the literature regarding BNs in healthcare. We looked for keywords such as "Bayesian Networks", "probabilistic graphical models", "medical", "clinical" to appear in the abstract of each paper. While it would not constitute a significant number, it is possible that some relevant papers were not included because they did not use the selected keywords in their abstract. This is especially true in cases where the actual name of the medical condition is described without mentioning the words "medical" or "clinical". However, we believe that the large number of selected papers was sufficient for drawing conclusions.

## 4. Summary

This paper began by recognising that while other authors had reviewed, and in some cases identified and quantified the medical conditions for which other AI and ML methods had been developed, in all cases they had eschewed BNs. This was our primary research problem. We also highlighted three secondary issues present in those reviews. That it was often difficult to identify: (a) the search terms they had used; (b) a framework or description for medical conditions; or (c) an accurate accounting of the papers identified by and used in each study.

As a core component of our systematic review of BNs in healthcare, our reviewers worked with clinicians to develop a list for use in identifying the target medical conditions that had been the focus of BN modelling. To address the primary research problem, we quantified the BNs from our literature collection by medical condition, identifying four conditions that received the majority of attention from authors and reviewing papers from each of those conditions in greater detail.

Also, to ensure we did not proliferate the three secondary issues observed of other works, we provided: (a) the search term and process used to arrive at our literature collection; (b) the list of medical conditions that reviewers were asked to use in classifying medical BNs; and (c) a URL link to the complete review dataset.

BNs have and continue to receive significant interest for their ability to combine available evidence and accurately reason under conditions of uncertainty. This paper has contributed to our understanding of how and what BNs have been considered for in healthcare. We identified the four primary conditions that have received almost two-thirds of BN-modelling attention in the literature: cardiac, cancer, psychological and lung disorders. We also found a number of acute differences in how BNs were being applied to each.

The wider observation is that despite the strong research interest in BN models in healthcare, this interest is not matched by adoption in practice. It is possible that with most BN in healthcare research effort going to these four primary conditions that already receive vast amounts of funding and attention from other solution areas, that BNs are being drowned out: effectively lost in the noise. It is also possible that a lack of understanding pervades with respect to how BNs work and what they are capable of. Either way, it is only through understanding and promotion, continued multidisciplinary research and adherence to the full disclosure required of the scientific method that we may in future harness the potential of BNs in daily healthcare practice and affect positive change, improving outcomes for clinicians and patients alike.

**Acknowledgement**
SM, EK, GAH and NF acknowledge support from the EPSRC under project EP/P009964/1: PAMBAYESIAN: Patient Managed decision-support using Bayes Networks. KD acknowledges financial support from Massey University for his study sabbatical and visits with the PamBayesian team.